%% file: ms.tex
\documentclass[12pt,preprint]{aastex}
\slugcomment{Submitted to Astron. J.}
\shortauthors{Haisch et al.}
\shorttitle{A Mid-Infrared Multiplicity Survey}
\begin{document}
\title{Mid-Infrared Observations of Class I/Flat-Spectrum Systems in Six Nearby Molecular Clouds}

\author{Karl E. Haisch Jr.\altaffilmark{1,2}}
\affil{Department of Physics, Utah Valley State College, 800 W. University Pkwy., Orem, Utah 84058-5999, haischka@uvsc.edu}

\and

\author{Mary Barsony\altaffilmark{1,2,3,4}}
\affil{Department of Physics \& Astronomy, San Francisco State University, 1600 Holloway Drive, San Francisco, California 94132, mbarsony@stars.sfsu.edu}

\and

\author{Michael E. Ressler\altaffilmark{1}}
\affil{Jet Propulsion Laboratory, Mail Stop 169-327, 4800 Oak Grove Drive, Pasadena, California 91109, Michael.E.Ressler@jpl.nasa.gov}

\and

\author{Thomas P. Greene}
\affil{NASA Ames Research Center, Mail Stop 245-6, Moffett Field, California  94035-1000, tgreene@mail.arc.nasa.gov}

\altaffiltext{1}{Observations with the Palomar 5m telescope were obtained under a collaborative agreement between Palomar Observatory and the Jet Propulsion Laboratory.}

\altaffiltext{2}{Visiting Astronomer at the Baade telescope of the Magellan Observatory, a joint facility of the Carnegie Observatories, Harvard University, Massachusetts Institute of Technology, University of Arizona, and the University of Michigan.}

\altaffiltext{3}{Space Science Institute, 4750 Walnut Street, Suite 205, Boulder, CO 80301}

\altaffiltext{4}{NASA Faculty Fellow}

\begin{abstract}

We have obtained new mid-infrared observations of 65 Class I/Flat-Spectrum (F.S.) objects in the Perseus, Taurus, Chamaeleon I and II, $\rho$ Ophiuchi, and Serpens dark clouds. These objects represent a subset of the young stellar objects (YSOs) from our previous near-infrared multiplicity surveys. We detected 45/48 (94\%) of the single sources, 16/16 (100\%) of the primary components, and  12/16 (75\%) of the secondary/triple components of the binary/multiple objects surveyed. One target, IRS 34, a 0$\farcs$31 separation F.S. binary, remains unresolved at near-infrared wavelengths.  The composite spectral energy distributions (SEDs) for all of our sample YSOs are either Class I or F.S., and, in 15/16 multiple systems, at least one of the individual components displays a Class I or F.S. spectral index.  However, the occurrence of mixed pairings, such as F.S. with Class I, F.S. with Class II, and, in one case, an F.S. with a Class III (Cha I T33B), is suprisingly frequent.
Such behavior is not consistent with that of multiple systems among  T Tauri stars (TTS), 
where the companion of a classical TTS also tends to be a classical TTS, although other mixed pairings have been previously observed among Class II YSOs. Based on an analysis of the spectral indices of the individual binary components, there appears to be a higher proportion of mixed Class I/Flat-Spectrum systems (65-80\%) than that of mixed Classical/Weak Lined T Tauri systems (25-40\%), demonstrating that the envelopes of Class I/Flat-Spectrum systems are rapidly evolving during this evolutionary phase. In general, the individual binary/multiple components suffer very similar extinctions, A$_{v}$, suggesting that most of the line-of-sight material is either in the foreground of the molecular cloud or circumbinary. We report the discovery of a steep spectral index secondary
companion to ISO-Cha I 97, detected for the first time via our mid-infrared observations.  In our
previous near-infrared imaging survey of binary/multiple Class I and F.S.YSOs, ISO-Cha I 97 appeared
to be single. With a spectral index of $\alpha$ $\geq$ $+$3.9, the secondary component of 
this system is a member of a rare class of very steep spectral index YSOs, those with $\alpha$ $>$ $+$3. Only three such objects have previously been reported, all of which are either Class 0 or Class I.

\end{abstract}


\section{Introduction}

We have known for many years that the majority of solar-to-late-type field stars are binaries or multiples (Abt \& Levy 1976; Duquennoy \& Mayor 1991; Fischer \& Marcy 1992), but only in the past decade have significant numbers of pre-main-sequence (PMS) stars been surveyed for multiplicity. Near-infrared surveys of nearby, young dark cloud complexes have shown that young, low-mass TTSs have binary fractions which are greater than or equal to that of the field (e.g., Ghez, Neugebauer, \& Matthews 1993; Mathieu 1994; Simon et al. 1995; Ghez et al. 1997; Barsony, Koresko, \& Matthews 2003), and appear to be coeval, that is, at the same evolutionary age (Hartigan, Strom, \& Strom 1994; Brandner \& Zinnecker 1997). Observations at millimeter continuum wavelengths also find multiple sources at the origins of extended molecular outflows and optical jets (Looney, Mundy, \& Welch 1997). Thus, the formation of binary and multiple systems appears to be the rule, rather than the exception, in star-forming regions.

Recently, multiplicity surveys have been extended to include even younger, self-embedded YSOs, those with Class I and flat-spectrum SEDs (Lada 1987; Adams, Lada, \& Shu 1987, 1988). The first survey for multiplicity among such objects was conducted by Looney, Mundy, \& Welch (2000) with the BIMA millimeter interferometer at 2.7 mm. Of the eight Class 0 and Class I YSOs surveyed, all were members of binary systems or small groups. Not long thereafter, Reipurth (2000) analyzed the multiplicity of 14 young (ages $\leq$ 10$^{5}$ yr) sources which drive giant Herbig-Haro flows. Between 79\% and 86\% (depending on the range of separations considered) of these sources had at least one companion and, of these, half were higher order multiple systems.  These fractions are even larger than those found among either the PMS TTS or field star populations. 

In order to determine the multiplicity properties of embedded Class I and flat-spectrum sources with reasonable statistical confidence, which the previous surveys do not provide, Haisch et al. (2002, 2004) and Duch\^{e}ne et al. (2004) conducted the first near-infrared imaging surveys of nearby molecular clouds which permitted observations at resolutions $\leq$ 1$^\prime$$^\prime$, and therefore enabled detections of companions as close as 100 - 300 AU, depending on the distance to the clouds. Merging the three surveys into one, these authors derived a ``restricted" companion star fraction ($\Delta$K $\leq$ 4 magnitudes and a separation range of 300 - 1400 AU) of 16\% $\pm$ 3\%, with all clouds presenting fully consistent fractions. This frequency is in excellent agreement with that obtained for TTSs in the same star-forming regions, and is approximately twice as high as that observed for late-type field dwarfs.

While near-infrared data are very efficient in identifying binary/multiple components, such observations are not at a long enough wavelength to determine the evolutionary state of the objects via their spectral indices (as defined in Section 3). However, since contamination from photospheric emission is minimal at mid-infrared wavelengths, observations at 10 $\mu$m, in conjunction with near-infrared $JHKL$ data, can be used to ascertain the spectral index, and thus the evolutionary state of the individual binary/multiple YSOs. The 10 $\mu$m radiation from each component does not originate from the star itself, but from a ``photosphere" of surrounding dust heated to several hundred degrees. Radiative transfer models of Class I YSOs have shown that the 10 $\mu$m photosphere is located about 1 AU from a low-mass protostar (e.g., Kenyon, Calvet, \& Hartmann 1993; Chick \& Cassen 1997). Thus, the individual components should be point sources in the mid-infrared. Indeed, recent 10 $\mu$m observations of Class I objects have often found them to be unresolved (e.g., Barsony, Ressler, \& Marsh 2005). Finally, no current space-based observatory can match the angular resolution achievable from the ground at 
mid-infrared wavelengths, which is ideal for studying very young binaries. 

Mid-infrared studies of small samples of Class I/flat-spectrum binaries have already been published (Liu et al. 1996; Girart et al. 2003; Ciardi et al. 2003). In this paper, we present the results of a mid-infrared imaging survey of 65 single and binary/multiple Class I and flat-spectrum YSOs from our previous near-IR surveys (Haisch et al. 2002; 2004) in the $\rho$ Ophiuchus ($d$ = 125 pc; Knude \& H{\o}g 1998), Serpens ($d$ = 310 pc; de Lara, Chavarria-K., \& L\'{o}pez-Molina 1991), Taurus ($d$ = 140 pc; Kenyon \& Hartmann 1995), Perseus ($d$ = 320 pc; Herbig 1998), and Chamaeleon I and II ($d$ = 160 pc and $d$ = 178 pc; Whittet et al. 1997) star-forming regions, respectively. Following the near-infrared multiplicity survey of Haisch et al. (2004), we impose a lower limit for detectable separations of 100 -- 140 AU for Taurus, Chamaeleon, and $\rho$ Ophiuchi and $\sim$ 300 AU for Perseus and Serpens, which was set by the seeing in the near-infrared. We have also imposed an upper limit to the separations of 2000 AU in order to avoid including sources which are not gravitationally bound systems (e.g., Reipurth \& Zinnecker 1993; Simon et al. 1995). In addition, sensitivity calculations from our near-infrared data indicate that we can detect a $K$ = 4 magnitude difference between the primary and companion at a separation of 1\arcsec \hspace*{0.05in}at the 5$\sigma$ confidence level. Thus, we restrict our analysis to component magnitude differences $\Delta$$K$ $\leq$ 4 mag.  We discuss our mid-infrared observations and data reduction procedures in $\S$2. In $\S$3, we present the results of our survey, and discuss the results in $\S$4. We summarize our primary results in $\S$5.

\section{Observations and Data Reduction}

The mid-infrared imaging data for this survey were obtained with MIRLIN, JPL's 128 $\times$ 128 pixel Si:As camera (Ressler et al. 1994), and MIRAC3/BLINC, a mid-infrared array camera built for astronomical imaging at the Steward Observatory, University of Arizona, and the Harvard Smithsonian Center for Astrophysics (Hoffmann et al. 1998). MIRAC3/BLINC utilizes a Rockwell HF-16 128 $\times$ 128 Si:As  BIB hybrid array. All observations at Palomar were made with MIRLIN, whereas MIRAC3/BLINC was used on the Baade 6.5m telescope.

Mid-infrared $N$-band ($\lambda_{o}$ = 10.78 $\mu$m, $\Delta$$\lambda$ = 5.7 $\mu$m) observations of all Taurus and Perseus sources were obtained with MIRLIN on 17 December 2003. Observations at $N$-band ($\lambda_{o}$ = 10.3 $\mu$m, $\Delta$$\lambda$ = 1.03 $\mu$m) of all sources in Chamaeleon I, and L1689 SNO2 and IRS 67 in $\rho$ Oph, were obtained with MIRAC3/BLINC during the period 17 - 19 March 2003. Fluxes for the remaining $\rho$ Oph sources surveyed were taken from Barsony, Ressler, \& Marsh (2005). MIRLIN has a plate scale of 0$\farcs$15 pix$^{-1}$ and a 19$\farcs$2 $\times$ 19$\farcs$2 field of view at the Palomar 5m telescope. At the Baade 6.5m telescope, MIRAC3/BLINC has a plate scale of 0$\farcs$123 pix$^{-1}$ and a 15$\farcs$7 $\times$ 15$\farcs$7 field of view. For reference, the full-width at half-maximum of a diffraction-limited image at $N$-band with MIRLIN is 0$\farcs$47 at the Palomar 5m telescope; the corresponding value with MIRAC3/BLINC is 0$\farcs$40 at the Baade 6.5m telescope.

Standard mid-IR chop-nod techniques were used for the data acquisition. For all MIRLIN and MIRAC3/BLINC observations, the telescope's secondary mirror was chopped 8$^\prime$$^\prime$ in a north-south direction, at a rate of a few Hz. In order to remove residual differences in the background level, the entire telescope was nodded 8$^\prime$$^\prime$ east-west. Several hundred coadded chop pairs, with 5 - 6 msec integration times per frame, were combined to produce images with a total on-source integration time of 25 sec with MIRLIN and 60 sec with MIRAC3/BLINC.

All MIRAC3/BLINC data were reduced using the Image Reduction and Analysis Facility (IRAF)\footnote[5]{IRAF is distributed by the National Optical Astronomy Observatories, which are operated by the Association of Universities for Research in Astronomy, Inc., under cooperative agreement with the National Science Foundation.}. All target frames were background-subtracted, shifted, and coadded to produce the final images of each object. For the MIRLIN data, all raw images were background-subtracted, shifted, and coadded with our in-house IDL routine, ``mac'' (match-and-combine).

Flux calibration was performed using $\alpha$ Tau ($N$ = -3.02) as our standard star for the MIRLIN observations, and both $\alpha$ CMa ($N$ = -1.35) and $\gamma$ Cru ($N$ = -3.36) for the MIRAC3/BLINC data. The standards were observed on the same nights and through the same range of airmasses as the target sources. Zero points and extinction coefficients were established for each night using a straight-line fit to the instrumental minus true magnitudes of the standards as a function of airmass. Typical airmass corrections for the MIRLIN data were of order 0.5 mags/airmass, while the corrections for the MIRAC3/BLINC data were of order 0.1 - 0.2 mags/airmass. By adding the errors in the zero-point offsets, the airmass corrections, the aperture corrections, and the magnitudes of the standards in quadrature and taking the square root, we estimate that the total photometric uncertainty for the MIRLIN observations is typically good to within $\pm$ 0.10 magnitudes, and $\pm$ 0.06 for the MIRAC3/BLINC data. Typical 1$\sigma$ 10 $\mu$m rms sensitivities were 0.040 Jy with both MIRLIN at Palomar and with MIRAC at Magellan. Note that 0.00 magnitudes at $N$-band with MIRLIN corresponds to 33.4 Jy, whereas 0.00 magnitudes with MIRAC3/BLINC corresponds to 35.2 Jy. 

Aperture photometry was performed for the MIRLIN data using 14 pixel (2$\farcs$1) radii for standards and 8 pixel (1$\farcs$2) radii for program objects. The corresponding apertures used for the MIRAC3/BLINC photometry had 10 pixel (1$\farcs$2) and 6 pixel (0$\farcs$7) radii for the standards and program objects, respectively. Sky values around each source were determined from the mode of intensities in an annulus with a radius of 3 pixels (0$\farcs$36 and 0$\farcs$45 for MIRAC and MIRLIN, respectively). Our choice of aperture size for our target photometry insured that the individual source fluxes were not contaminated by the flux from companion stars, however they are not large enough to include all the flux from a given source. In order to account for this missing flux, aperture corrections were derived from the flux standards, and the instrumental magnitudes for all sources were corrected to account for the missing flux.

\section{Analysis and Results}

We detected 45/48 (94\%) of the single sources, 16/16 (100\%) of the primary components, and 12/16 (75\%) of the secondary/triple components of the near-infrared binary/multiple objects in our mid-infrared survey. One system, IRS34, is resolved as a binary in the mid-infrared, but remains unresolved at near-infrared wavelengths.  The mid-infrared photometry for our target objects is listed in Tables~\ref{table1} --~\ref{table5}. The first column in each table lists the common name for each source, followed by the objects' J2000 coordinates in the second and third columns respectively. The fourth column lists the objects' 10 $\mu$m flux, or flux upper limit, in Jy units. The last two columns list the UT observation date, and the telescope/instrument combination used. 

Spectral energy distributions were constructed for the binary/multiple objects in Table~\ref{table1} using the tabulated 10 $\mu$m fluxes and $JHKL$ magnitudes from Haisch et al. (2002, 2004) converted to fluxes in Jy units. Each source was classified using the least squares fit to the slope between 2.2 and 10 $\mu$m. We calculated the spectral indices from 2.2 to 10 $\mu$m for all observed sources from the relation:

\begin{equation}
\alpha = \frac {d log(\lambda F_\lambda)}{d log \lambda}
\end{equation}

\noindent in order to quantify the natures of their SEDs (Lada 1987). Our calculated spectral indices are typically good to $\alpha$ = $\pm$ 0.3, as discussed in the next section. Table \ref{table6} lists the 2.2 -- 10 $\mu$m spectral indices, $\alpha$, for all sources. The classification scheme of Greene et al. (1994) has been adopted in our analysis as it is believed to correspond well to the physical stages of evolution of YSOs (e.g. Andr\'{e} \& Montmerle 1994). Class I sources have $\alpha$ $>$ 0.3, flat spectrum sources have 0.3 $>$ $\alpha$ $\geq$ --0.3, Class II sources have --0.3 $>$ $\alpha$ $\geq$ --1.6, and sources with $\alpha$ $<$ --1.6 are Class III YSOs. A Class I object is thought to be one in which the central YSO has attained essentially its entire initial main-sequence mass, but is still surrounded by a remnant infall envelope and an accretion disk. Flat-spectrum YSOs represent a transitional class between Class I and Class II objects (e.g. Kikuchi, Nakamoto, \& Ogochi 2002). They are characterized by near-infrared spectra that are strongly veiled by continuum emission from hot, circumstellar dust. Class II sources are surrounded by accretion disks, while Class III YSOs have remnant, or absent, accretion disks. Thus, in this context, the progression from the very red Class I YSO {\bf $\rightarrow$} flat spectrum {\bf $\rightarrow$} Class II {\bf $\rightarrow$} Class III has frequently been interpreted as representing an evolutionary sequence, even though Reipurth (2000) suggested that more violent transition from the embedded to the optically bright stages could occur when components are ejected from unstable multiple systems.

In Figure~\ref{figure1}, we present the $JHKL$ color-color diagram for each individual component
of the binary/multiple systems detected in our mid-infrared survey, excepting ChaI T33B and Ced 110 IRS6, for which we do not have $L$-band data, and IRS34, which is unresolved shortwards of 10 $\mu$m. All sources are plotted showing their SED classifications. Class I sources are designated with a pentagon, flat spectrum sources with a square, and Class II sources with a star. In the diagram, we plot the locus of points corresponding to both the unreddened main sequence (MS) and giant branch (Bessell \& Brett 1988) as a solid line. The Classical T Tauri star (CTTS) locus (Meyer, Calvet, \& Hillenbrand 1997) is shown as a dot-dashed line. The two leftmost parallel dashed lines define the reddening band for main sequence stars and are parallel to the reddening vector. The length of the arrow above these lines corresponds to the displacement produced by 10 magnitudes of visual extinction. The reddening law of Cohen et al. (1982) has been adopted.  All but four of the Class I and flat spectrum sources lie in the infrared excess region of the $JHKL$ color-color diagram. The four exceptions are: GY 263, both components of the IRS 51 system, and the secondary of IRS 54. In addition, only one of the Class II sources, GY 23, lies in the infrared excess region of the color-color diagram.

In Figure~\ref{figure2} we show the variation of $K - L$ with $K - N$ for the same objects as in Figure~\ref{figure1}. Sources are plotted with their SED classifications using the same symbols as in Figure~\ref{figure1}. The leftmost vertical dashed line represents the $K - N$ color for a source with $\alpha$ = -0.3 ($K - N$ = 4.5), and the rightmost vertical dashed line represents the $K - N$ color for a source with $\alpha$ = $+$0.3 ($K - N$ = 5.4). Thus, Class I sources should lie to the right of the rightmost vertical dashed line, flat spectrum sources should lie between the two vertical dashed lines, and Class II sources should lie to the left of the leftmost vertical dashed line. The length of the arrow in the diagram corresponds to the displacement produced by 10 magnitudes of visual extinction, and has its origin at the photospheric colors of an M4 star ($K - L$ = 0.29, $K - N$ = 0.13; Bessell \& Brett 1988, Ducati et al. 2001). 

Visual extinction estimates, A$_{v}$, listed in Table~\ref{table6} for all except the Class I sources, were determined by dereddening each source in the $JHKL$ color-color diagram using the extinction law of Cohen et al. (1982). Complications due to heating and reprocessing of both stellar and disk radiation in the remnant infall envelopes preclude accurate dereddening for the Class I sources via this method.  For sources in the infrared excess region of the $JHKL$ color-color diagram, we dereddened each source to the CTTS locus. For sources to the right of the termination point of the CTTS locus, we used adopted intrinsic colors for the Class II and Flat Spectrum sources.  For the sources in the reddening band of Figure \ref{figure1}, median intrinsic colors of ($J - H$)$_{o}$ = 0.62 and ($H - K$)$_{o}$ = 0.1 were adopted (Strom, Strom, \& Merrill 1993). For the Class II sources beyond the termination point of the CTTS locus, we adopted intrinsic colors of ($J - H$)$_{o}$ = 0.8 and ($H - K$)$_{o}$ = 0.5, while for flat spectrum sources ($H - K$)$_{o}$ = 0.75 was used (Strom, Strom, \& Merrill 1993; Greene \& Meyer 1995). 

\section{Discussion}

\subsection{SED Characteristics}

The composite SEDs for all of our sample YSOs are either Class I or Flat-Spectrum. However, individual source components frequently display Class II, or in one case Class III, spectral indices.  In most cases, the SED classes of the primary and secondary components are different, a Class I object being paired with a Flat-Spectrum source, or a Flat-Spectrum source paired with a Class II YSO. In one instance (ChaI T33B) we also find a Flat-Spectrum source paired with a Class III object, and in another (EC92/EC95) a Class I/Class II pairing. Such behavior is not consistent with what one typically finds for TTSs, where the companion of a classical TTS also tends to be a classical TTS (Prato \& Simon 1997; Duch\^{e}ne et al. 1999). Mixed pairings, however, have been observed among Class II YSOs previously (Ressler \& Barsony 2001; McCabe et al. 2006). 

We plot the primary's spectral index vs. the secondary's spectral index for our Class I/Flat-Spectrum sample  in Figure 3(a). For comparison, we show the same plot in Figure 3(b) for Class II YSOs from data in Barsony et al. (2005) and McCabe et al. (2006). For a given primary spectral index, there is a larger spread in the spectral indices of the secondaries among the Class I/Flat-Spectrum sample compared to the Class II sample. Thus, there appears to be a higher proportion of mixed Class I/Flat-Spectrum systems (79\%) than that of mixed Classical/Weak-Lined T Tauri systems (25\%) (Hartigan \& Kenyon 2003; Prato, Greene, \& Simon 2003; Barsony, Ressler, \& Marsh 2005; McCabe et al. 2006). Allowing for the $\alpha$ = $\pm$ 0.3 error in our calculated spectral indices, we find the proportion of mixed Class I/Flat-Spectrum systems (65\%) remains higher than that of mixed Classical/Weak-Lined T Tauri systems (42\%). Given the low likelihood of misclassification of Class I and Class II YSOs as discussed below, this demonstrates that the envelopes of Class I/Flat-Spectrum systems are rapidly evolving during this evolutionary phase, although they may still be coeval. Although the central objects of Class I and Flat-Spectrum sources (Doppmann et al 2005; Covey et al. 2005), and even Class II YSOs (Kenyon et al. 1998; White \& Hillenbrand 2004), have similar stellar and accretion properties, their envelope properties clearly differ.

In general, the individual components of a given binary/multiple system suffer very similar extinctions, A$_{v}$, suggesting that most of the line-of-sight material is either in the foreground of the molecular cloud or circumbinary. One notable exception is the GY 244/WL 5 pair, whose A$_{v}$ values differ by {\it at least} 11 magnitudes. While their projected separation is $\sim$ 1100 AU (Haisch et al. 2004), it is possible that GY 244 and WL 5 represent a chance projection.  WL 5 is known to be one of the most heavily extincted infrared sources in the $\rho$ Ophiuchi cloud core (Andr\'{e} et al. 1992), consistent with our measured value of A$_{v}$ $\geq$ 44.0. Our results are consistent with those of previous authors, who find
WL5 to be a Class III YSO seen through large amounts of foreground cloud material (Andr\'{e} \& Montmerle 1994 and references therein; Bontemps et al. 2001).

The superiority of using the $KLN$ color-color diagram to specify the SED
classes of YSO's relative to the traditionally used $JHKL$ color-color diagram 
is evident upon comparison of Figures 1 \& 2. In the $KLN$ color-color diagram
of Figure 2, the Class II, Flat Spectrum, and Class I objects are very cleanly
separated, whereas they remain relatively confused in the $JHKL$ color-color diagram 
of Figure 1.  Furthermore, Class II sources cannot be distinguished from
reddened background stars in Figure 1 (all but one lie in the reddening band), 
whereas they all lie in a clearly demarcated region in Figure 2.

Source variability may affect the SED classification of an object in the sense
that, in general, the near- and mid-infrared fluxes for a given object in this survey
were not acquired simultaneously. To estimate the possible effect
of source variability on SED class, we note that YSO variability in the K-band has
been observed at amplitudes ranging from 0.32 to 0.65 mag in Ophiuchus (Barsony et al. 1997),
from 0.15 to 2.2 mag in Serpens, with typical amplitudes $<$ 0.75 mag (Kaas 1999), and with
typical amplitudes of 0.2 mag in Orion A (Carpenter et al. 2001). For most of our source sample,
the L-band magnitudes presented in Table 6 are the only ones available, so we cannot
estimate typical L-band variations one might observe in the YSO populations. Our L-band 
photometric errors are typically 0.1-0.2 magnitudes for the fainter sources. 
In a recent large-scale mid-infrared survey of Ophiuchus YSOs, at least 20\% of all embedded sources are known to have significant flux variations, ranging from 0.2 to 1.8 mag at N band, with typical amplitudes $<$ 0.5 mag (Barsony, Ressler, \& Marsh 2005). Using the typical variations in magnitude at K- and N-band, if a source varies in such a fashion that it becomes bluer, i.e., increasing 2 $\mu$m emission and decreasing 10 $\mu$m emission, alpha could vary by -0.3. If a source varies in the sense of becoming redder, that is, its 2 $\mu$m flux is decreasing as its 10 $\mu$m flux increases, then alpha could vary by $+$0.3. These are worst case effects on misclassifying an object's SED class.

Referring to Figure 1, there are two Class I and two F.S. sources 
that lie blueward of the reddening band.  The position of these 
sources in the $JHKL$ color-color diagram cannot be explained using
spherically symmetric radiative transfer models for the embedded YSOs.
Two-dimensional radiative transfer models, however, can reproduce the 
location of these YSOs in Figure 1, through a combination of the presence
of outflow cavities carved out of the infall or remnant infall envelopes and 
inclination effects (Whitney et al. 2003a,b). The two Class I sources which lie to the left of  the reddening band in Figure~\ref{figure1}
are the secondary components of IRS 51 ($J - H$ = $>$ 7.20, $K - L$ = 2.01) and 
of IRS 54 ($J - H$ = $>$ 3.87, $K - L$ = 1.15). IRS 51 has been examined in detail by Haisch  et al. (2002), who concluded that the mid-infrared secondary may be a bright knot in the molecular outflow from IRS 51 (Bontemps et al. 1996), along a North-South cavity seen in scattered light at $K$-band, rather than a true YSO. We note, however, that Duchene et al. (2004) consider the secondary to IRS 51 to be a real source as it appears indistinguishable from a point source in their image of the system. The secondary to IRS 54 was undetected in our survey.  Its flux upper limit 
is consistent with either a Class I  or later SED class YSO ($\alpha$ $\leq$ $+$1.49). Two flat-spectrum sources, the primary of IRS 51 ($J - H$ = 6.56, $K - L$ = 1.81) and GY 263 ($J - H$ = $>$ 5.10, $K - L$ = 1.71), also lie to the left of the reddening band Figure~\ref{figure1}. With a spectral index of $\alpha$ = -0.29, the primary of IRS 51 is right at the borderline between Flat-Spectrum and Class II. Its designation as a Flat-Spectrum source is supported by Barsony, Ressler, \& Marsh (2005), who derive a spectral index of $+$0.13 for the IRS 51 primary. In the case of GY 263, Barsony, Ressler, \& Marsh (2005) derive a spectral index of $\alpha$ = -0.4 for the source, making it a Class II YSO. 

Whereas the sources discussed above lie in regions of the $JHKL$ color-color diagram 
that obviate corrrect determination of their SED classes, 
in the $KLN$ color-color diagram of Figure~\ref{figure2},
they all lie at their expected locations .  In Figure 2, there is a clear progression from the very red Class I YSO {\bf $\rightarrow$} Flat Spectrum {\bf $\rightarrow$} Class II. 

While near-to-mid-infrared colors generally reflect the evolutionary state of a given object, we note that there may be instances in which the SED class does not yield the correct evolutionary state. For example, since the flux at 2 $\mu$m suffers heavier extinction than the 10 $\mu$m flux, a Class II YSO seen through a large amount of foreground obscuration may show a Class I or Flat-spectrum SED. Orientation effects may also be important, since a nearly pole-on Class I source may display an SED similar to an edge-on Class II object (Whitney et al. 2003b). Nevertheless, protostellar envelopes extinct the stellar flux over a much larger range of solid angle than disks. 
Assuming a random distribution of orientations in 2-D, and the most likely orientation to be
45$^{\circ}$, if we assume an inclination angle in the range 45$^{\circ}\le\,i\,\le135^{\circ}$, 
then 70\% of Class I YSOs will be properly classified.
Furthermore, if one assumes a random distribution, in 3-D, of disk orientations,
the probability of observing a disk edge-on (85$^{\circ}\le\,i\,\le95^{\circ}$) is 9\%
Thus, only 9\% of Class II sources would be improperly
classified as Class I (B. Whitney, priv. comm.) Furthermore, the SEDs of Class I and Class II YSOs
do look distinctly different, with Class I sources exhibiting larger far-infrared excesses.

The rigorously correct way to determine an object's evolutionary state is to obtain multiwavelength imaging data for each source, and quantitatively compare these data to models produced using 3D radiative transfer codes (Whitney et al. 2003b). For example, Osorio et al. (2003) have self-consistently modeled each component of the Class I binary L1551 IRS 5, adopting a flattened infalling envelope surrounding a circumbinary disk. A wealth of observations for this system provide additional constraints for any viable source models: ISO imaging, spectroscopy of the water ice feature with SpeX, and high-resolution VLA imaging at 7 mm resolving the binary disks. The results show that a flattened-envelope collapse model is required to explain simultaneously the large-scale fluxes and the water ice and silicate features. The circumstellar disks are optically thick in the millimeter range and are inclined so that their outer parts hide the emission along the line of sight from their inner parts. Furthermore, these disks have lower mass accretion rates than the infall rate of the envelope. Unfortunately, such multiwavelength data are not available for our sources, thus we must rely on the spectral indices as an indicator of YSO evolutionary states.

\subsection{Notes on Selected Sources}

{\bf 03260$+$3111}: This source was observed to be binary in the near-infrared, with a 3\farcs62 separation and position angle of 47.9 degrees (Haisch et al. 2004). In the present survey only the K-band primary is detected. The system appears extended with a fan-shaped nebulosity in the mid-infrared. Based on an optical and near-infrared imaging survey, Magnier et al. (1999) have classified 03260$+$3111 as a flat-spectrum YSO. Our mid-infrared photometry suggests that the primary is a Class II object, while our upper limit for the flux of the secondary suggests that it is at least a Class II YSO.

{\bf Ced 110 IRS 6}: The primary of this binary system (separation = 1$\farcs$95, P.A. = 95.6$^{\circ}$) has the largest visual extinction (A$_{v}$ $\geq$ 51.0) in our survey. With a spectral index of $+$0.30, it is right on the borderline between Flat-Spectrum and Class I YSOs, and, therefore, may, in fact, be a Class I object. If so, Ced 110 IRS 6 would represent the only binary in our survey in which both components are Class I. Mid- and far-infrared observations with ISOCAM and ISOPHOT suggest that most of the flux density is associated with the primary component of the binary (Lehtinen et al. 2001).

{\bf ISO-Cha I 97}: ISO-Cha I 97 was detected as a single star in our near-infrared imaging survey of binary/multiple Class I and Flat-Spectrum YSOs (Haisch et al. 2004). Our mid-infrared observations have revealed that this source is in fact binary (separation = 2$\farcs$05, P.A. = 72.6$^{\circ}$). Our 5$\sigma$ $K$-band sensitivity limit of $\sim$ 18.5 combined with our 10 $\mu$m flux yields a lower limit to the spectral index of the secondary component of ISO-Cha I 97 of $\alpha$ $\geq$ $+$3.9 (the primary component has a spectral index $\alpha$ = $+$0.50). This very steep spectral index places the secondary of ISO-Cha I 97 in a class of YSO that has heretofore been rarely known, i.e., those with $\alpha$ $>$ $+$3. Three such objects have been recently reported, the Class 0 object Cep E mm (Noriega-Crespo et al. 2004), source X$_{E}$ in R CrA (Hamaguchi et al. 2005), and source L1448 IRS 3(A) (Ciardi et al. 2003; Tsujimoto, Kobayashi, \& Tsuboi 2005). Further very steep spectrum YSOs are expected to be discovered with the Spitzer Space Telescope.

{\bf IRS 34}: As is the case for ISO-Cha I 97, IRS 34 was observed to be a single source in the near-infrared (Haisch et al. 2004), but binary in the mid-infrared (Barsony, Ressler, \& Marsh 2005). The components of IRS 34 are separated by only 0$\farcs$31 (P.A. 236$^{\circ}$), and therefore could not be resolved in our near-infrared survey. For this reason, we are also unable to construct SEDs for the individual components of IRS 34.

Finally, we note that we did not detect the secondaries of the Cha I T33B, IRS 48, IRS 54, and GY 51 binary/multiple sources in the mid-infrared. We derived upper limits for the SED classifications for these components based on our mid-infrared flux upper limits. These classifications are likely rather firm for the secondaries of Cha I T33B and IRS 48, however given the large extinctions observed toward IRS 54 (A$_{v}$ $>$ 25 mag; Haisch et al. 2004) and GY 51 (A$_{v}$ $\simeq$ 30), the SED classifications for these components are less secure. The secondary of IRS 54 appears to be a Class I YSO. Accounting for the extinction toward GY 51 (see Duchene et al. 2004 for a K-band image of this triple system), the components of this system would become Class II-III, Class III, Class II for the primary, secondary, and third components respectively.

\section{Summary and Conclusions}

We have obtained new mid-infrared observations of 64 Class I/flat-spectrum objects in the Perseus, Taurus, Chamaeleon I and II, $\rho$ Ophiuchi, and Serpens dark clouds. These objects represent a subset of the YSOs from our previous near-infrared multiplicity surveys (Haisch et al. 2002, 2004). We detected 45/48 (94\%) of the single sources, 16/16 (100\%) of the primary components, and 12/16 (75\%) of the secondary/triple components of the binary/multiple objects surveyed. The tight mid-IR binary, IRS34, remains unresolved at near-IR wavelengths.

While the composite SEDs for all of our sample YSOs are either Class I or Flat-Spectrum, individual source components may display Class II, or, in one case, Class III, spectral indices.  The SED classes of the primary and secondary components are frequently different. For example, a Class I object may be found to be paired with a Flat-Spectrum source, or a Flat-Spectrum source paired with a Class II YSO. Such behavior is not consistent with what one typically finds for TTSs, where the companion of a classical TTS also tends to be a classical TTS (Prato \& Simon 1997; Duch\'{e}ne et al. 1999). Mixed pairings, however, have been previously observed among Class II YSOs (Ressler \& Barsony 2001; McCabe et al. 2006).

Based on an analysis of the spectral indices of the individual binary components, there appears to be a higher proportion of mixed Class I/Flat-Spectrum systems (65-80\%) than that of mixed Classical/Weak Lined T Tauri systems (25-40\%). This demonstrates that the envelopes of Class I/Flat-Spectrum systems are rapidly evolving during this evolutionary phase, although they may still be coeval.

In general, the individual binary/multiple components suffer very similar extinctions, A$_{v}$, suggesting that most of the line-of-sight material is either in the foreground of the molecular cloud or circumbinary. However, the GY 244/WL 5 pair, whose A$_{v}$ values differ by 11 magnitudes, is a notable exception.  If the projected separation of this pair ($\sim$ 1100 AU), equivalent to its physical separation, then this system could easily be gravitationally bound. However, it is also possible that GY 244 and WL 5 represent a chance projection.

ISO-Cha I 97 was detected as a single star in our near-infrared imaging survey of binary/multiple Class I and Flat-Spectrum YSOs (Haisch et al. 2004). Our mid-infrared observations have revealed that this source is in fact binary. Combining our $K$-band sensitivity limit from HGBS04 with our 10 $\mu$m flux yields a lower limit to the spectral index of the secondary component of ISO-Cha I 97 of $\alpha$ $\geq$ $+$3.9. This very steep spectral index places the secondary of ISO-Cha I 97 in a class of YSO that has heretofore been rarely known, i.e., those with $\alpha$ $>$ $+$3. Further very steep spectrum YSOs are expected to be discovered with the Spitzer Space Telescope.

\acknowledgements

We thank the Magellan Observatory staff for their outstanding support in making our observations possible. K. E. H. gratefully acknowledges partial support from NASA Origins grant NAG 5-11905 to Ray Jayawardhana. M. B. acknowledges NSF grant AST-0206146 and NASA LTSA Grant NAG5-8933 to
Barbara Whitney, which made her contributions to this work possible, as well as a NASA Summer Faculty Fellowship held at NASA's Ames Research Center during the summers of 2003-2004. T. P. G. acknowledges support from NASA Origins of Solar Systems grant 344-37-22-11.

\newpage

\clearpage
\input{tab1.tex}
\clearpage
\input{tab2.tex}
\clearpage
\input{tab3.tex}
\clearpage
\input{tab4.tex}
\clearpage
\input{tab5.tex}
\clearpage
\input{tab6.tex}
\clearpage

\figcaption[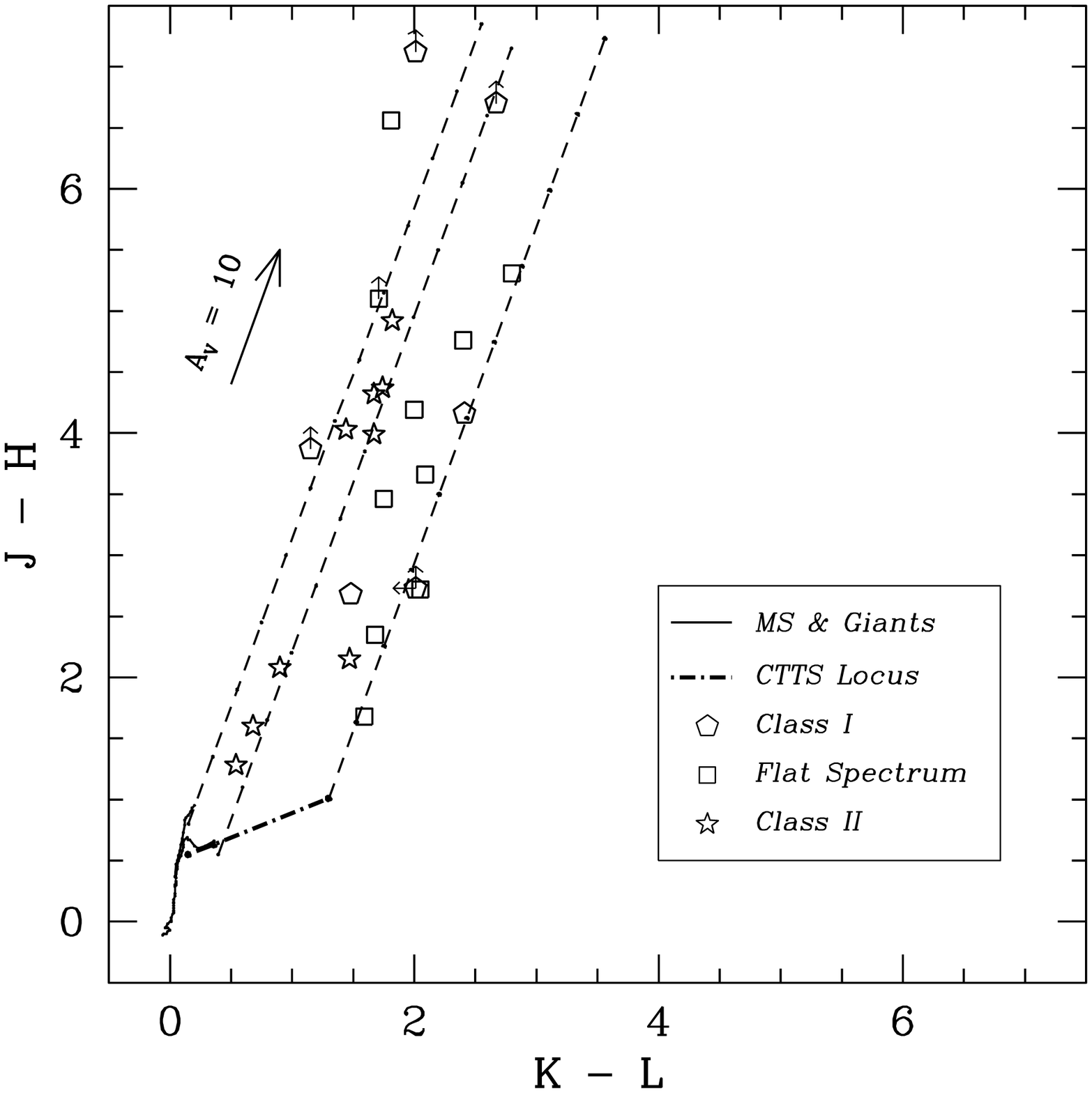]
{$JHKL$ color-color diagram for all sources in our mid-infrared binary/multiple survey. All sources are plotted showing their SED classifications. Class I sources are designated with a pentagon, flat-spectrum sources with a square, and Class II sources with a star. We plot the locus of points corresponding to both the unreddened main-sequence (MS) and giant branch as a solid line, and the classical T Tauri star locus as a dot-dashed line. The two leftmost parallel dashed lines define the reddening band for main-sequence stars and are parallel to the reddening vector. The length of the arrow above these lines corresponds to the displacement produced by 10 magnitudes of visual extinction. The rightmost dashed line is parallel to the reddening vector, and has its origin at the colors of the reddest T Tauri star. \label{figure1}
}

\figcaption[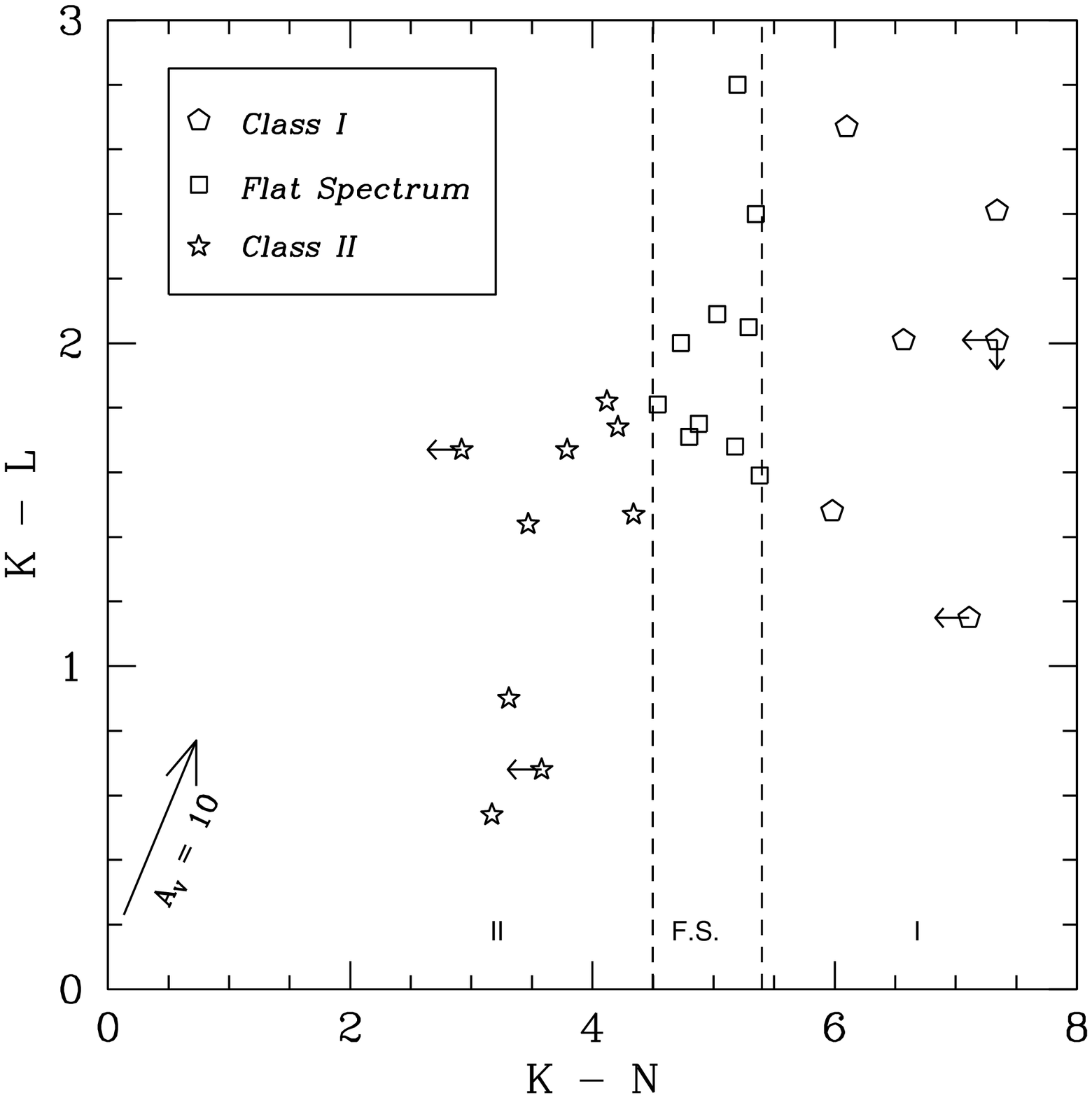]
{Color-color diagram showing the variation of $K - L$ with $K - N$ for all sources in Figure 1; the same symbols are used. The area demarcated by the vertical dashed lines indicates the realm of the Flat Spectrum sources:  The leftmost vertical dashed line represents the $K - N$ color for a source with $\alpha$ = -0.3 ($K - N$ = 4.5), and the rightmost vertical dashed line represents the $K - N$ color for a source with $\alpha$ = $+$0.3 ($K - N$ = 5.4). The length of the arrow in the diagram corresponds to the displacement produced by 10 magnitudes of visual extinction, and has its origin at the photospheric colors of an M4 star ($K - L$ = 0.29, $K - N$ = 0.13).  Arrows on selected objects indicate upper limits on the $K - L$ or $K - N$ colors.  \label{figure2}
}

\figcaption[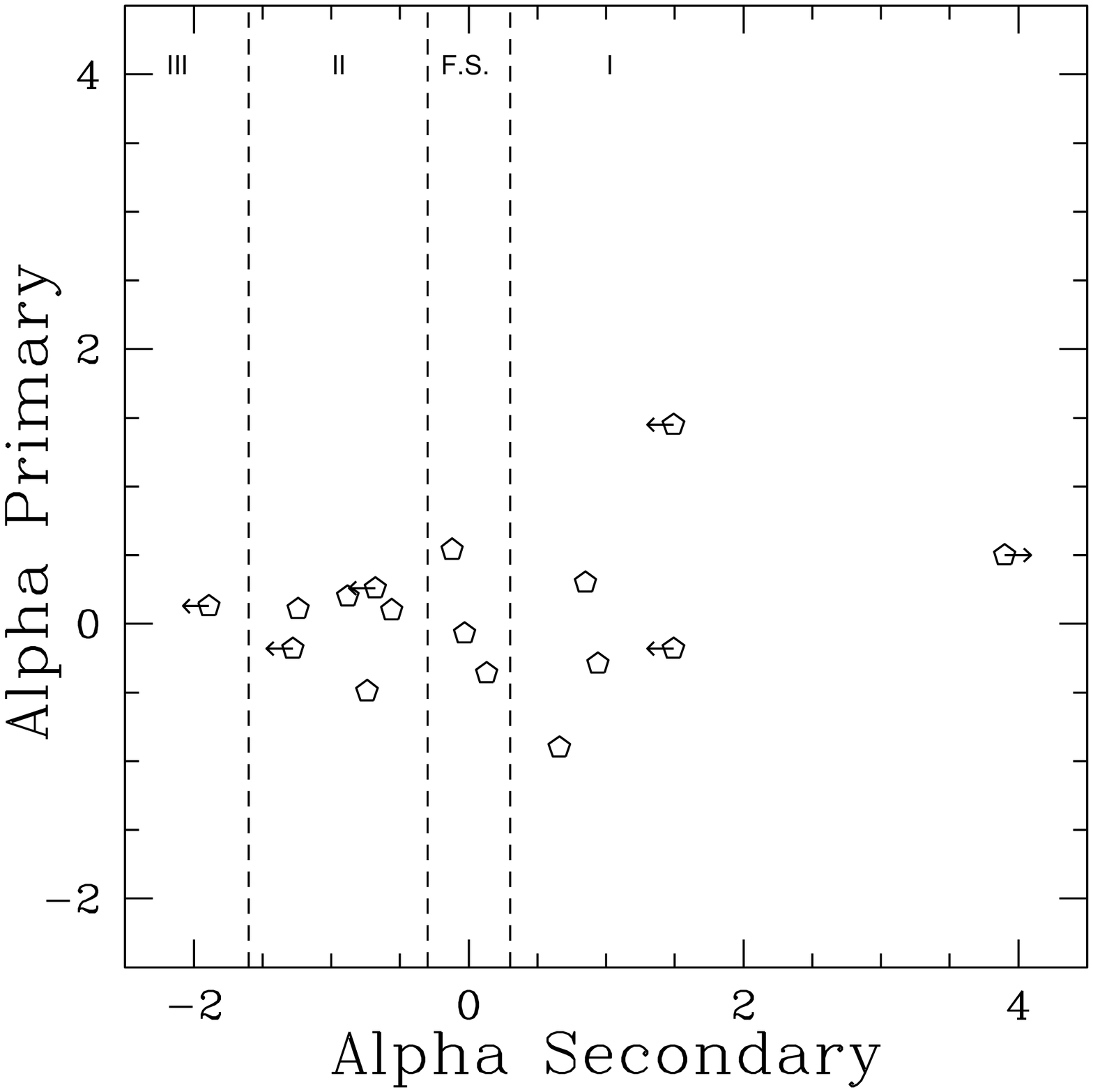]
{Figure 3(a):  Plot of the primary's spectral index vs. the secondary's spectral index for our Class I/Flat-Spectrum sample.  Figure 3(b): Plot of the same quantity for a sample of Class II YSOs from Barsony, Ressler, \& Marsh (2005) and  McCabe et al. (2006).  \label{figure3a}
}

\clearpage
\plotone{f1.eps}
\clearpage
\plotone{f2.eps}
\clearpage
\plotone{f3a.eps}
\clearpage
\plotone{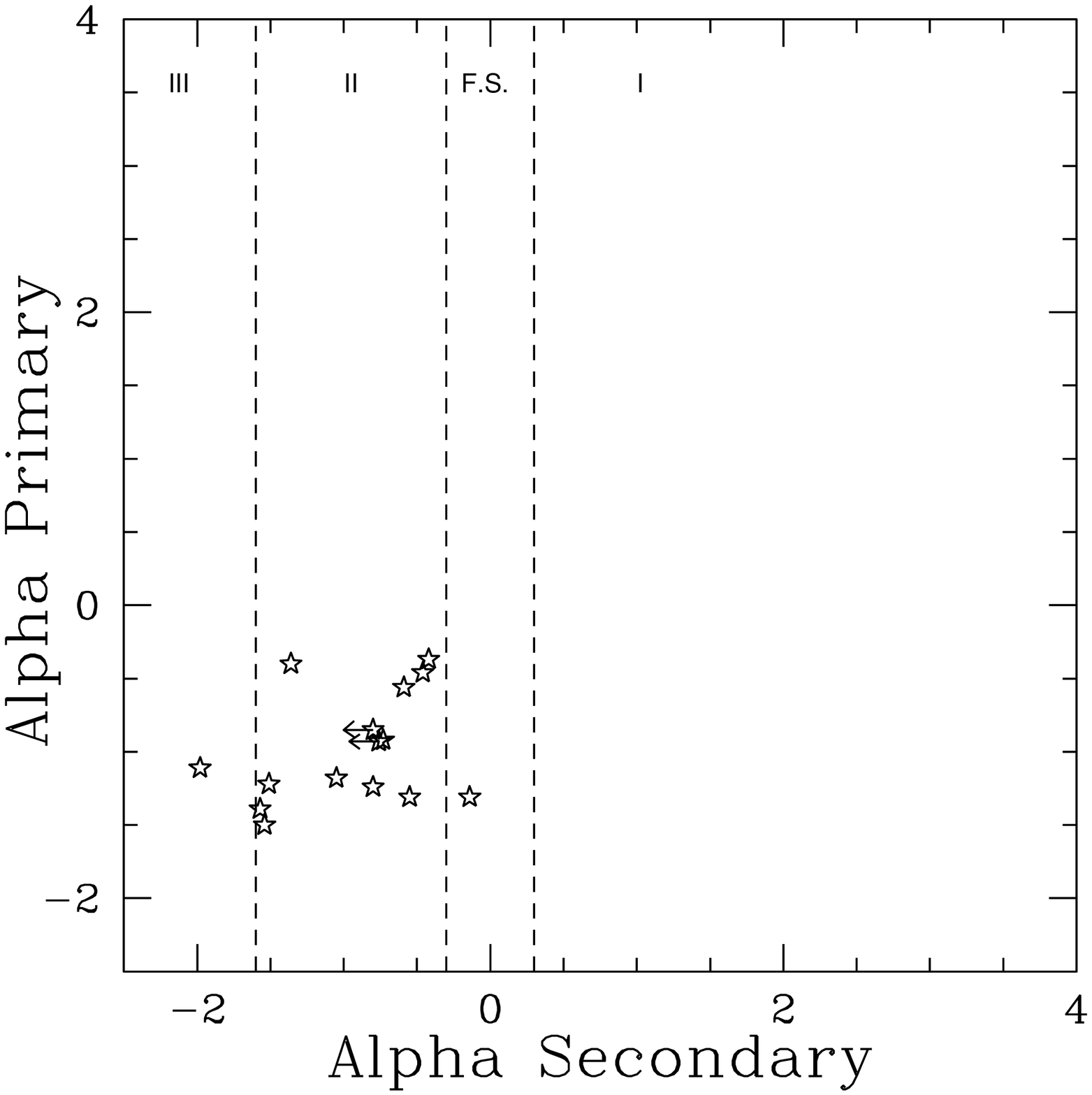}
\clearpage

\end{document}

%% file: tab1.tex
\begin{deluxetable}{lccrcc}
\footnotesize
\tablecaption{Mid-Infrared 10 $\mu$m Fluxes for Multiple Sources
\label{table1}}
\tablewidth{0pt}
\tablehead{Source & RA(J2000) & Dec(J2000) & F$_{10 \mu m}$ (Jy) & Obs. Date & Tel./Inst.\tablenotemark{a} }
\startdata
03260$+$3111\tablenotemark{b} & 03 29 10.40 & $+$31 21 58.0 & 0.751        & 17Dec03 & P200/MIRLIN\\
 secondary                                   &  03 29 10.61  &  $+$31 22 00.43   &$\leq$0.12 & 17Dec03 & P200/MIRLIN\\
 & & & & & \\  
ChaI T33B                                    & 11 08 15.69 & -77 33 47.1      & 5.96          & 17Mar03 & M6.5m/MIRAC \\
 secondary                                   &  11 08 14.98  &  -77 33 46.5 & $\leq$ 0.045 & 17Mar03 & M6.5m/MIRAC \\
 & & & & & \\ 
Ced110 IRS6 & 11 07 09.80 & -77 23 04.4 & 0.210 & 17Mar03 & M6.5m/MIRAC \\
     secondary &  11 07 10.40  &  -77 23 04.6  & 0.077 & 17Mar03 & M6.5m/MIRAC \\
 & & & & & \\
ISO-ChaI 97 & 11 07 18.30 & -77 23 13.0 & 0.210 & 17Mar03 & M6.5m/MIRAC\\
   secondary &  11 07 18.90  &  -77 23 13.6  & 0.048 & 17Mar03 & M6.5m/MIRAC\\
 & & & & & \\
GY 23 & 16 26 24.00 & -24 24 49.9 & 2.02 & 02Jul98 & P200/MIRLIN\\
GY 21 & 16 26 23.54 & -24 24 41.5 & 0.40 & 02Jul98 & P200/MIRLIN\\
 & & & & & \\
IRS 48 & 16 27 37.20 & -24 30 34.0 & 3.91 & 30Jun98 & P200/MIRLIN\\
IRS 50 & 16 27 38.10 & -24 30 40.0 & $\leq$ 0.097 & 01Jul98 & P200/MIRLIN\\
 & & & & & \\   
L1689 SNO2 & 16 31 52.13 & -24 56 15.2 & 1.95 & 17Mar03 & M6.5m/MIRAC\\
 secondary    &  16 31 51.94  &  -24 56 13.7  & 0.213 & 17Mar03 & M6.5m/MIRAC\\
 & & & & & \\
IRS 51 & 16 27 39.84 & -24 43 16.1 & 0.730 & 25Jun97 & P200/MIRLIN\\
 secondary & 16 27 39.86  &   -24 43 17.5 & 0.480 & 25Jun97 & P200/MIRLIN\\
 & & & & & \\
IRS 43 & 16 27 26.90 & -24 40 51.5 & 1.54 & 07Jun98 & KeckII/MIRLIN\\
 secondary &  16 27 26.88 &  -24 40 51.1  & 0.52 &  07Jun98 & KeckII/MIRLIN\\
GY 263 & 16 27 26.60 & -24 40 45.9 & 0.030 & 07Jun98 & KeckII/MIRLIN\\
 & & & & & \\
IRS 54\tablenotemark{c} & 16 27 51.70 & -24 31 46.0 & 2.51 & 26Jun97 & P200/MIRLIN\\
 secondary                    &   16 27 51.39 &  -24 31 40.1  & $\leq$ 0.045 & 26Jun97 & P200/MIRLIN\\
 & & & & & \\ 
GY 51 & 16 26 30.49 & -24 22 59.0 & 0.224 & 07Jun98 & KeckII/MIRLIN\\
 secondary &  16 26 30.57  &  -24 22 59.5  & $\leq$ 0.018 & 07Jun98 & KeckII/MIRLIN\\
 tertiary      & 16 26 30.90 &  -24 22 58.2  & $\leq$ 0.018 & 07Jun98 & KeckII/MIRLIN\\
 & & & & & \\ 
WL 1 & 16 27 04.13 & -24 28 30.7 & 0.08 & 07Jun98 & KeckII/MIRLIN\\
 secondary & 16 27 04.09 & -24 28 30.1 & 0.05 & 07Jun98 & KeckII/MIRLIN\\
 & & & & & \\
WL 2         & 16 26 48.56 & -24 28 40.4 & 0.139 & 07Jun98 & KeckII/MIRLIN\\
 secondary & 16 26 48.47 & -24 28 36.4 & 0.016 & 07Jun98 & KeckII/MIRLIN\\
 & & & & & \\
GY 244 & 16 27 17.54 & -24 28 56.5 & 0.194 & 24Apr96 & P200/MIRLIN\\
WL 5 & 16 27 18.00 & -24 28 55.0 & 0.050\tablenotemark{d} & 28Jun99 & KeckI/LWS\\
 & & & & & \\
IRS 34\tablenotemark{e} & 16 27 15.48 & -24 26 40.6 & 0.092 & 07Jun98 & KeckII/MIRLIN\\
 secondary                      &  16 27 15.46  &  -24 26 40.8  & 0.080 & 07Jun98 & KeckII/MIRLIN\\
 & & & & & \\ 
SVS 20 & 18 29 57.70 & $+$01 14 07.0 & 4.36 & 07Jun98 & KeckII/MIRLIN\\
 secondary & 18 29 57.72 & $+$01 14 08.5 & 1.53 & 07Jun98 & KeckII/MIRLIN\\
 & & & & & \\
EC 95 & 18 29 57.80 & $+$01 12 52.0 & 0.100 & 07Jun98 & KeckII/MIRLIN\\
EC 92 & 18 29 57.75 & $+$01 12 57.0 & 0.550 & 07Jun98 & KeckII/MIRLIN\\
\enddata
\tablenotetext{a}{P200/MIRLIN, Keck II/MIRLIN, and KeckI/LWS data for all objects listed in this table are from Barsony, Ressler, \& Marsh (2005).}
\tablenotetext{b}{03260$+$3111 was resolved in the near-IR with a 3$\farcs$26 separation at PA 47.9$^{\circ}$ (Haisch et al. 2004). The source is unresolved, but extended with a fan-shaped nebulosity in the mid-IR.}
\tablenotetext{c}{IRS 54 is a binary source in the near-IR, but the secondary was not detected in the mid-IR.}
\tablenotetext{d}{Flux listed for WL 5 is at 12.5 $\mu$m.}
\tablenotetext{e}{IRS 34 was not resolved in the near-IR due to the tight separation (0\farcs31) and faintness of the individual components.}
\end{deluxetable} 

%% file: tab2.tex
\begin{deluxetable}{lccrcc}
\footnotesize
\tablecaption{Mid-Infrared 10 $\mu$m Fluxes for Single Perseus Sources
\label{table2}}
\tablewidth{0pt}
\tablehead{Source & RA(J2000) & Dec(J2000) & F$_{10 \mu m}$ (Jy) & Obs. Date & Tel./Inst.}
\startdata
03382$+$3145 & 03 41 22.70 & $+$31 54 46.0 & $\leq$ 0.031 & 17Dec03 & P200/MIRLIN\\
03259$+$3105 & 03 29 03.70 & $+$31 15 52.0 & 7.24 & 17Dec03 & P200/MIRLIN\\ 
03262$+$3114 & 03 29 20.40 & $+$31 24 47.0 & $\leq$ 0.024 & 17Dec03 & P200/MIRLIN\\
03380$+$3135 & 03 41 09.10 & $+$31 44 38.0 & 0.244 & 17Dec03 & P200/MIRLIN\\ 
03220$+$3035 & 03 25 09.20 & $+$30 46 21.0 & 0.328 & 17Dec03 & P200/MIRLIN\\
03254$+$3050 & 03 28 35.10 & $+$31 00 51.0 & 0.219 & 17Dec03 & P200/MIRLIN\\
03445$+$3242 & 03 47 41.60 & $+$32 51 43.5 & 1.05 & 17Dec03 & P200/MIRLIN\\ 
03439$+$3233 & 03 47 05.00 & $+$32 43 09.0 & 0.199 & 17Dec03 & P200/MIRLIN\\
\enddata
\end{deluxetable}

%% file: tab3.tex
\begin{deluxetable}{lccrcc}
\footnotesize
\tablecaption{Mid-Infrared 10 $\mu$m Fluxes for Single Taurus Sources
\label{table3}}
\tablewidth{0pt}
\tablehead{Source & RA(J2000) & Dec(J2000) & F$_{10 \mu m}$ (Jy) & Obs. Date & Tel./Inst.}
\startdata
Haro 6-13 & 04 32 15.61 & $+$24 29 02.3 & 1.39 & 17Dec03 & P200/MIRLIN\\
Haro 6-28 & 04 35 55.87 & $+$22 54 35.5 & 0.089 & 17Dec03 & P200/MIRLIN\\
04489$+$3042 & 04 52 06.90 & $+$30 47 17.0 & 0.278 & 17Dec03 & P200/MIRLIN\\ 
04016$+$2610 & 04 04 42.85 & $+$26 18 56.3 & 2.46 & 17Dec03 & P200/MIRLIN\\
04108$+$2803A\tablenotemark{a} & 04 13 52.90 & $+$28 11 23.0 & 0.839 & 17Dec03 & P200/MIRLIN\\
04361$+$2547 & 04 39 13.87 & $+$25 53 20.6 & 0.724 & 17Dec03 & P200/MIRLIN\\ 
04365$+$2535 & 04 39 35.01 & $+$25 41 45.5 & 0.972 & 17Dec03 & P200/MIRLIN\\
04295$+$2251 & 04 32 32.10 & $+$22 57 30.0 & 0.391 & 17Dec03 & P200/MIRLIN\\
04264$+$2433 & 04 29 30.30 & $+$24 39 54.0 & 0.432 & 17Dec03 & P200/MIRLIN\\
\enddata
\tablenotetext{a}{The Class I object, IRAS 04108$+$2803B ($\alpha_{2000}=$04$^h$13$^m$54.69$^s$, $\delta_{2000}=$28$^{\circ}$11$^{\prime}$33\farcs1), at 21$^{\prime\prime}$ separation from 04108$+$2803A,
falls outside our 2000 AU separation limit, but the pair have been considered to be a binary system by previous authors (e.g.,
Barsony \& Kenyon 1992).}
\end{deluxetable}

%% file: tab4.tex
\begin{deluxetable}{lccrcc}
\footnotesize
\tablecaption{Mid-Infrared 10$\mu$m Fluxes for Single Chamaeleon Sources
\label{table4}}
\tablewidth{0pt}
\tablehead{Source & RA(J2000) & Dec(J2000) & F$_{10 \mu m}$ (Jy) & Obs. Date & Tel./Inst.}
\startdata
ChaI T32 & 11 08 04.61 & -77 39 16.9 & 2.99 & 17Mar03 & M6.5m/MIRAC\\
ChaI T44 & 11 10 01.35 & -76 34 55.8 & 3.88 & 17Mar03 & M6.5m/MIRAC\\
ChaI T42 & 11 09 54 66 & -76 34 23.7 & 5.82 & 17Mar03 & M6.5m/MIRAC\\
ChaI T29 & 11 07 59.25 & -77 38 43.9 & 3.66 & 17Mar03 & M6.5m/MIRAC\\
ISO-ChaI 26 & 11 08 04.00 & -77 38 42.0 & 0.790 & 19Mar03 & M6.5m/MIRAC\\
ChaI C1-6 & 11 09 23.30 & -76 34 36.2 & 0.910 & 17Mar03 & M6.5m/MIRAC\\
ChaI C9-2 & 11 08 37.37 & -77 43 53.5 & 4.01 & 17Mar03 & M6.5m/MIRAC\\
\enddata
\end{deluxetable}

%% file: tab5.tex
\begin{deluxetable}{lccrcr}
\footnotesize
\tablecaption{Mid-Infrared 10$\mu$m Fluxes for Single $\rho$ Oph Sources
\label{table5}}
\tablewidth{0pt}
\tablehead{Source & RA(J2000) & Dec(J2000) & F$_{10 \mu m}$ (Jy)\tablenotemark{a} & Obs. Date & Tel./Inst.}
\startdata
Elias 29 & 16 27 09.43 & -24 37 18.5 & 23.7 & 26Jun97 & P200/MIRLIN\\
IRS 42 & 16 27 21.45 & -24 41 42.8 & 2.03 & 27Jun97 & P200/MIRLIN\\ 
GSS 30/IRS 1 & 16 26 21.50 & -24 23 07.0 & 8.17 & 30Jun98 & P200/MIRLIN\\
GY 279 & 16 27 30.18 & -24 27 44.3 & 0.954 & 27Jun97 & P200/MIRLIN\\
IRS 63 & 16 31 35.53 & -24 01 28.3 & 0.948 & 17Mar03 & M6.5m/MIRAC\\
GSS 26 & 16 26 10.28 & -24 20 56.6 & 1.01 & 27Jun97 & P200/MIRLIN\\ 
WL 17 & 16 27 06.79 & -24 38 14.6 & 0.825\tablenotemark{b} & 29Jun99 & KeckI/LWS\\
IRS 67 & 16 32 01.00 & -24 56 44.0 & 0.822 & 17Mar03 & M6.5m/MIRAC\\
WL 12 & 16 26 44.30 & -24 34 47.5 & 1.14 & 01Jul98 & P200/MIRLIN\\
IRS 44 & 16 27 28.00 & -24 39 34.3 & 2.65 & 27Jun97 & P200/MIRLIN\\
IRS 46 & 16 27 29.70 & -24 39 16.0 & 0.540 & 07Jun98 & KeckII/MIRLIN\\
WL 6 & 16 27 21.83 & -24 29 53.2 & 1.01 & 07Jun98 & KeckII/MIRLIN\\
GY 224 & 16 27 11.17 & -24 40 46.7 & 0.255 & 07Jun98 & KeckII/MIRLIN\\
WL 19 & 16 27 11.74 & -24 38 32.1 & 0.130 & 24Apr96 & P200/MIRLIN\\
WL 3 & 16 27 19.30 & -24 28 45.0 & 0.162 & 27Jun97 & P200/MIRLIN\\
CRBR 15 & 16 26 19.30 & -24 24 16.0 & 0.074\tablenotemark{b} & 28Jun99 & KeckI/LWS\\
CRBR 12 & 16 26 17.30 & -24 23 49.0 & 0.315 & 27Jun97 & P200/MIRLIN\\
GY 344 & 16 27 45.81 & -24 44 53.7 & 0.117\tablenotemark{b} & 28Jun99 & KeckI/LWS\\
IRS 33 & 16 27 14.60 & -24 26 55.0 & 0.062 & 07Jun98 & KeckII/MIRLIN\\
GY 245 & 16 27 18.50 & -24 39 15.0 & 0.068 & 07Jun98 & KeckII/MIRLIN\\
CRBR 85 & 16 27 24.68 & -24 41 03.7 & 0.216 & 27Jun97 & P200/MIRLIN\\
GY 91 & 16 26 40.60 & -24 27 16.0 & 0.157 & 27Jun97 & P200/MIRLIN\\
WL 22 & 16 26 59.30 & -24 35 01.0 & 0.443 & 30Jun98 & P200/MIRLIN\\
GY 197 & 16 27 05.40 & -24 36 31.0 & $\leq$ 0.042 & 30Jun98 & P200/MIRLIN\\
\enddata
\tablenotetext{a}{MIRLIN and LWS fluxes taken from Barsony, Ressler, \& Marsh (2005).}
\tablenotetext{b}{The fluxes listed for WL 17, CRBR 15, and GY 344 are 12.5 $\mu$m fluxes.}
\end{deluxetable}

%% file: tab6.tex
\begin{deluxetable}{lrrrrrrcr}
\footnotesize
\tablecaption{Colors, Spectral Indices and Extinction Estimates for Multiple Sources
\label{table6}}
\tablewidth{0pt}
\tablehead{Source & $K$ & $(J - H)$ & $(H - K)$ & $(K - L)$ & $(K - N)$ & $\alpha$ & SED & A$_{v}$}
\startdata
03260$+$3111     & 7.29      & 1.28     & 0.79        & 0.54   & 3.17   & -0.91 & II      & 6.5 \\
                                & 11.04     & 1.54    & 1.03       & 0.82 & $<$ 3.44  & $<$ -0.84 & II - III & 6.8 \\
  & & & \\
ChaI T33B & 6.93 & 1.22 & 1.17 & --- & 1.93 & $+$0.13 & F.S. & 2.6 \\
 & 8.85 & 0.82 & 0.31 & --- & $<$ 1.62 & $\leq$ -1.89 & III & 3.2 \\
 & & & \\ 
Ced110 IRS6 & 10.86 & $>$ 4.82 & 4.32 & --- & 5.30 & $+$0.30 & F.S. & $\geq$ 51.0 \\
 & 12.86 & $>$ 2.43 & 4.71 & --- & 6.21 & 0.85 & I & ... \\
 & & & \\
ISO-ChaI 97 & 11.20 & --- & --- & --- & 5.64 & $+$0.50 & I &  ... \\
 & --- & --- & --- & --- & --- & $\geq$ $+$3.90 & 0(?) & ... \\
 & & & \\
GY 23 & 7.39 & 2.15 & 1.20 & 1.47 & 4.34 & -0.36 & II & 11.0 \\
GY 21 & 10.09 & 2.72 & 1.78 & 2.05 & 5.29 & $+$0.13 & F.S. & 16.0 \\
 & & & \\
IRS 48 & 7.71 & 1.68 & 1.01 & 1.59 & 5.38 & $+$0.26 & F.S. & 6.8 \\
IRS 50 & 9.92 & 1.60 & 1.02 & 0.68 & $<$ 3.58 & $\leq$ -0.68 &  II & 8.5\\
 & & & \\   
L1689 SNO2 & 8.32 & 2.35 & 1.49 & 1.68 & 5.18 & $+$0.20 & F.S. & 12.0 \\
 & 8.86 & 2.08 & 1.19 & 0.90 & 3.31 & -0.88 & II & 14.0 \\
 & & & \\
IRS 51 & 8.69 & 6.56 & 2.35 & 1.81 & 4.54 & -0.29 & F.S. & 34.0 \\
 & 11.18 & $>$ 7.12 & 1.47 & 2.01 & 6.57 & $+$0.94 & I & ... \\
 & & & \\
IRS 43 & 9.44 & $>$ 6.70 & 3.12 & 2.67 & 6.10 & $+$0.54 & I & ... \\
GY 263 & 12.42 & $>$ 5.10 & 1.77 & 1.71 & 4.80 & -0.12 & F.S. & $\geq$ 38.0 \\
 & & & \\
IRS 54 & 10.15 & 4.16 & 2.07 & 2.41 & 7.34 & $+$1.45 & I & ... \\
 & 14.29 & $>$ 3.87 & 1.84 & 1.15 & $<$ 7.11 & $\leq$ $+$1.49 & I & ... \\
 & & & \\ 
GY 51 & 10.16 & 4.19 & 2.06 & 2.00 & 4.73 & -0.18 & F.S. & 30.0 \\
 & 11.09 & 3.99 & 1.90 & 1.67 & $<$ 2.92  & $\leq$ -1.28 & II & 31.0 \\
 & 15.51 & $>$ 2.73 & 1.76 & $<$ 2.01 & $<$ 7.34 & $\leq$  $+$1.49 & I & ... \\
 & & & \\ 
WL 1\tablenotemark{a} & 10.76 & 4.37 & 2.08 & 1.74 & 4.21 & -0.49 & II & 35.0 \\
 & 10.85 & 4.32 & 2.06 & 1.67 & 3.79 & -0.74 & II & 33.0 \\
 & & & \\
WL 2 & 11.15 & 5.31 & 2.80 & 2.80 & 5.20 & $+$0.10 & F.S. & 38.0 \\
 & 12.42 & 4.92 & 2.73 & 1.82 & 4.12 & -0.56 & II & 38.0 \\
 & & & \\
GY 244 & 10.94 & 4.76 & 3.52 & 2.40 & 5.35 & $+$0.11 & F.S. & 33.0 \\
WL 5 & 10.21 & $>$ 5.47 & 4.82 & 2.27 & $<$ 2.59 & $<$-1.61 & III & $\geq$ 44.0 \\
 & & & \\
SVS 20 & 7.09 & 3.46 & 1.65 & 1.75 & 4.88 & -0.07 & F.S. & 24.0 \\
 & 8.38 & 3.66 & 1.92 & 2.09 & 5.03 & -0.03 & F.S. & 22.0  \\
 & & & \\
EC 95 & 9.78 & 4.03 & 2.12 & 1.44 & 3.47 & -0.90 & II & 32.0 \\
EC 92 & 10.44 & 2.68 & 1.59 & 1.48 & 5.98 & $+$0.66& I & ... \\
\enddata
\tablenotetext{a}{WL1 has an F.S. spectral index in Bontemps et al. (2001), and was resolved into binary Class II components by  Barsony, Ressler, \& Marsh (2005).}
\end{deluxetable}